\documentstyle[epsfig]{czjphys2}

\def\be{\begin{equation}}
\def\ee{\end{equation}}
\def\bea{\begin{eqnarray}}
\def\eea{\end{eqnarray}}
\newcommand{\la}{\langle}
\newcommand{\ra}{\rangle}
\begin{document}
\title{OLD AND NEW PARTON DISTRIBUTION AND FRAGMENTATION FUNCTIONS\footnote{
The work is supported by RFBR Grant No. 00-02-16696.}}
\authori{A.V. Efremov\,\footnote{E-mail: efremov@thsun1.jinr.ru}}
\addressi{Joint Institute for Nuclear Research, Dubna, 141980 Russia}
\authorii{}     
\authoriii{}     
\headtitle{Old and new PDF's and PFF's }
\headauthor{A.V. Efremov}  
\specialhead{A.V. Efremov: Old and new PDF's and PFF's.}
\evidence{A}  \daterec{XXX}    
\cislo{0}  \year{2001} \setcounter{page}{1} \pagesfromto{000--000}

\maketitle

\begin{abstract}
A short review of problems with parton distribution functions in nucleons,
non-polarized and polarized, is given. The main part is devoted to the
transversity distribution its possible measurement and its first
experimental probe via spin asymmetry in semi-inclusive DIS. It is argued
that the proton transversity distribution
could be successfully measured in future DIS experiments with
{\it longitudinally} polarized target.
\end{abstract}

\section{Parton characteristics of hadrons}
It is well known that three most important (twist-2) parton distributions
functions (PDF) in a nucleon are the non-polarized distribution function
$f_1(x)$, the longitudinal spin distribution  $g_1(x)$ and the transverse
spin distribution  $h_1(x)$ \cite{transversity}. The first two have
been more or less successfully measured experimentally in classical
deep inelastic scattering (DIS) experiments but the measurement of
the last one is especially difficult since it belongs to the class of
the so-called chiral-odd structure functions and can not be seen
there.

The non-polarized PDF's was measured for decades and are rather well
known in wide range of $x$ and $Q^2$. Its behavior in $Q^2$ is well
described by the QCD evolution equation (DGLAP) and serves as one of
the main source of $\alpha_s(Q^2)$ determination. The most recent
parametrization of these functions can be found in \cite{mrst98}.

One of the main problem here is the very small $x$ behavior. Summing
the leading $\log x$ with the help of the BFKL equation predicts a
rather quick rise of $xf_1(x)\propto x^{-0.55}$ which seems to find
some experimental support. Recent calculations of NLL corrections,
however, cardinally change this situation \cite{fadin98}.
Another problem is the flavor asymmetry of the sea quarks connected
with the break down of the Gottfried sum rule \cite{kumano98}.

The longitudinal spin PDF's draw common attention during last decade
in connection with the famous "Spin Crisis", i.e. astonishingly small
portion of the proton spin carried by quarks (see \cite{ael} and
references therein).  The most popular explanation of this phenomenon
is large contribution of the gluon spin $\Delta G(x)$.  The direct
check of this hypothesis is one of the main problem of future
dedicated experiments like COMPASS at CERN.  Even now, however, there
are some indication to considerable value of $\Delta G(x)$ coming
from the $Q^2$ evolution of the polarized PDF's \cite{leader99} with
the result $\int_0^1dx\Delta G(x)=0.58\pm0.12$ at $Q^2=1\,GeV^2$ and
from the first direct experimental probe of $\Delta G(x)$ by HERMES
collaboration \cite{hermes00} with the result $\Delta
G(x)/G(x)=0.41\pm0.18$ in the region $0.07<x<0.28$ (see Fig. 1).
The latter is in reasonable agreament with large $N_c$ limit prediction
\cite{efremov00} $\Delta G(x)/G(x)\approx 1/N_c$ for not very small $x$.

\begin{figure}[ht]
\begin{center}
\raisebox{-4mm}{
\mbox{\epsfig{figure=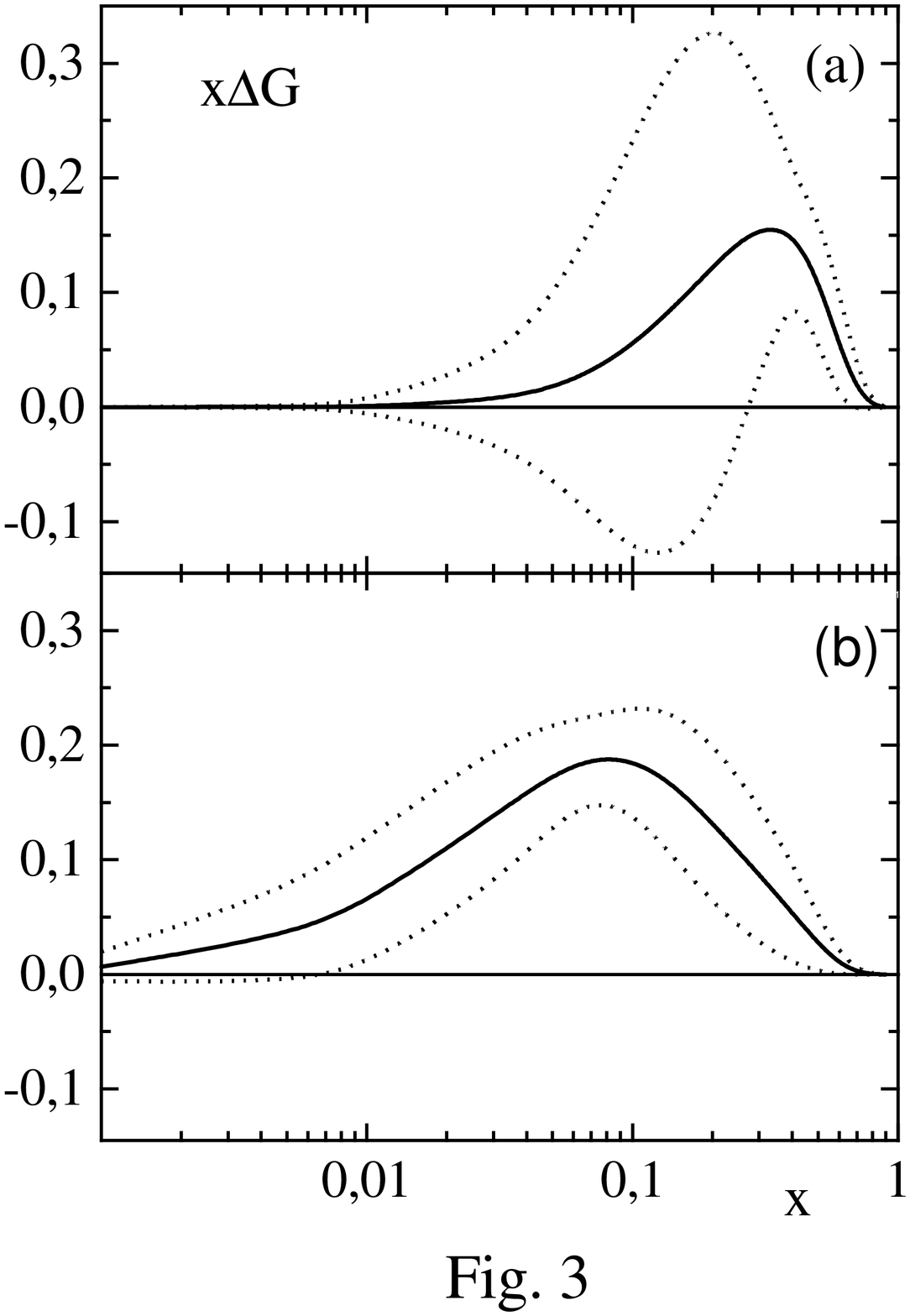,width=6.0cm,height=6.95cm}}}
\mbox{\epsfig{figure=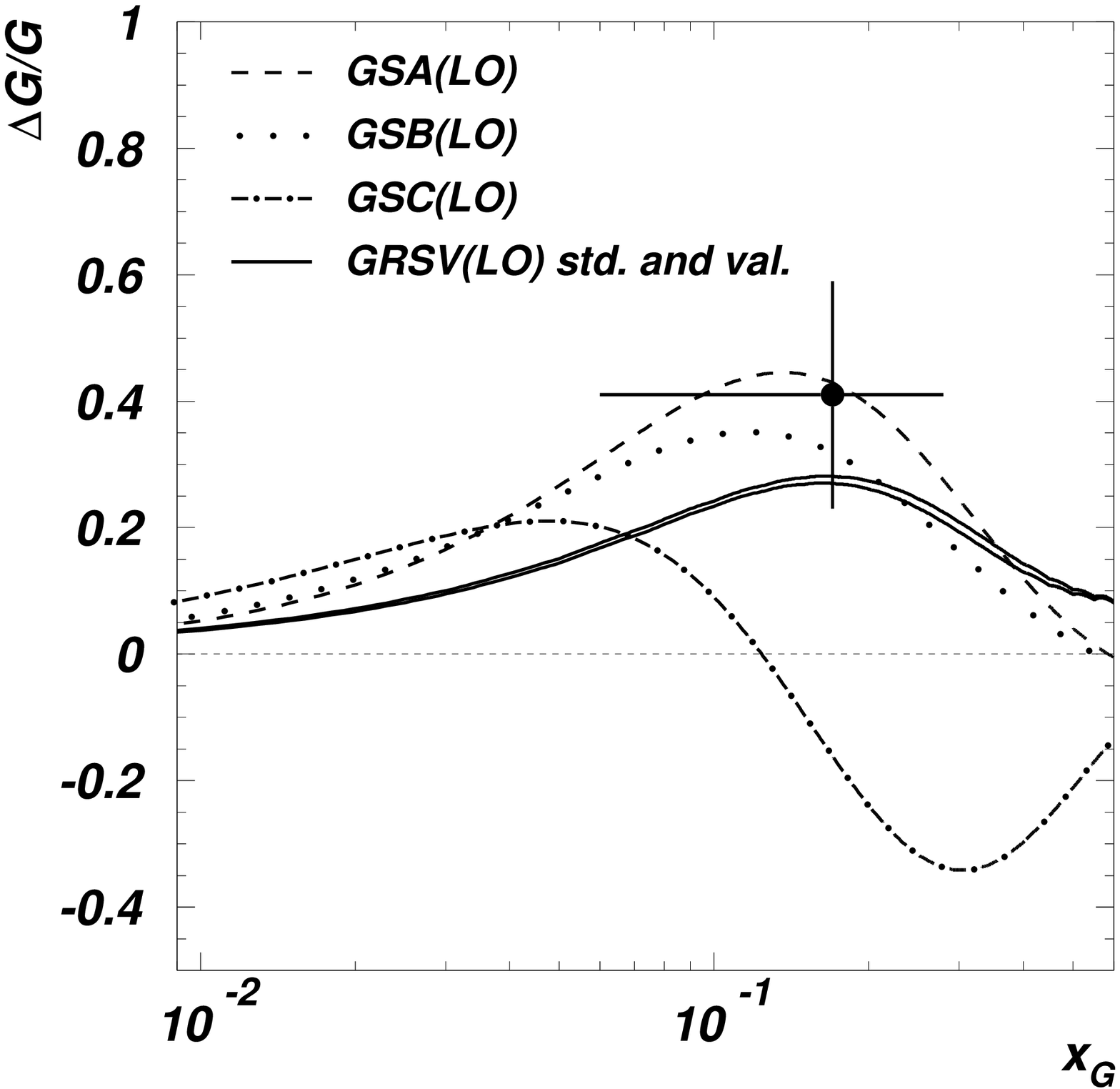,width=6.0cm,height=6.5cm}}
\end{center}
{\footnotesize
{\bf Fig. 1.} The gluon spin distributions obtained from ({\bf left})
fit of $Q^2$ evolution of polarized PDF's \cite{leader99}
( (a) without E155 data and (b) including E155)  and ({\bf right})
from high $p_T$ events in semi inclusive DIS \cite{hermes00}.
The curves present different fits of polarized DIS.}
\end{figure}

Another problem here is the sea quark spin asymmetry. It is usually assumed
in fitting the experimental data that $\Delta\bar u=\Delta\bar d
=\Delta\bar s$. This {\it ad hoc} assumption however contradicts with large
$N_c$ limit prediction $\Delta\bar u\approx-\Delta\bar d$ \cite{efremov00}.
This was previously discovered for the instanton model \cite{DK} and
supported by calculations in the chiral quark soliton model
\cite{DPPPW-96,Dressler00}.  Also this agrees with the phenomenological
approach based on "Pauly blocking principle" \cite{Gluck00}. An indication
to a nonzero value for $\Delta \bar{u} -\Delta \bar{d}$  was also observed
in \cite{MY}.

Turn now to DIS with transversely polarized target. A new result on the
structure function $g_2$ was recently reported by E155x collaboration
\cite{Rock00}. 
its twist-2 part is well defined by the structure function $g_1$ {\it via}
Wandzura-Wilczek relation \cite{WW77}
\be
g_2^{WW}(x,Q^2)=-g_1(x,Q^2)+ \int_x^1\frac{dy}{y}g_1(y,Q^2)
\ee
and the deflection off this is just pure dynamical twist-3 contribution.
This deflection have to be small however due to two exact sum rules
automatically valid for the WW-part: the Burkhardt-Cottingham \cite{BG70}
\be
\int_0^1dx g_1(y,Q^2)=0 \qquad \mbox{ (exper. $-0.01\pm0.02$) }
\ee
and Efremov-Leader-Teryaev \cite{ELT97}
\be
\int_0^1dxx\left(g_1^{val}(y,Q^2)+2g_2^{val}(x,Q^2)\right)=0 \qquad
\mbox{ (exper. $-0.004\pm0.008$) }
\ee

Concerning the transversity distribution it was completely unknown
experimentally till recent time. The only information comes from the Soffer
inequality \cite{soffer95} $|h_1(x)|\le{1\over2}[f_1(x)+ g_1(x)]$ which
follows from density matrix positivity. To access these chiral-odd
structure functions one needs either to scatter two polarized protons and
to measure the transversal spin correlation $A_{NN}$ in Drell-Yan process
(what is the problem for future RHIC) or to know the transverse
polarization of the quark scattered from transversely polarized target.
There are several ways to do this:

\begin{enumerate}
\item
To measure a polarization of a self-analyzing hadron to which the
quark fragments in a semi inclusive DIS (SIDIS), e.g.
$\Lambda$-hyperon \cite{augustin}.  The drawback of this method
however is a rather low rate of quark fragmentation into
$\Lambda$-particle ($\approx 2\%$) and especially that it is mostly
sensitive to $s$-quark polarization.
\item
To measure a transverse handedness in multi-particle parton
fragmentation \cite{hand}, i.e. the correlation of the quark spin
4-vector $s_\mu$ and particle momenta $k_i^\nu$,
$\epsilon_{\mu\nu\sigma\rho}s^\mu k_1^\nu k_2^\sigma k^\rho$
($k=k_1+k_2+k_3+\cdots$ is a jet 4-momentum).
\item
To use a new spin dependent T-odd parton fragmentation function (PFF)
\cite{muldt,muldz,mulddis} responsible for the left-right asymmetry
in one particle fragmentation of transversely polarized quark with
respect to quark momentum--spin plane. (The so-called "Collins
asymmetry" \cite{collins}.)
\end{enumerate}

The last two methods are comparatively new and only in the last years some
experimental indications to the transversal handedness and to the T-odd PFF
have appeared \cite{czjp99,todd}. The latter result was used \cite{egpu00}
to extract the information on the proton transversity distribution
\footnote{A similar work was done also in the paper \cite{aram} where some
adjustable parametrization for the T-odd PFF and some estimations for
$h_1(x)$ were used. Our approach is free of any adjustable parameters.}
from recently observed azimuthal asymmetry in SIDIS by HERMES \cite{hermes}
collaboration and by SMC \cite{bravardis99}. This will be the subject of the
rest of my talk.

\section{T-odd PFF's}

Analogous of PDF's $f_1,\ g_1$ and $h_1$ are the PFF's $D_1,\ G_1$ and
$H_1$, which describe the fragmentation of a non-polarized quark into
a non-polarized hadron  and a longitudinally or transversely
polarized quark into a longitudinally or transversely polarized
hadron, respectively.

These PFF's are integrated over the transverse
momentum ${\bm P}_{h\perp}$ of a hadron with respect to a quark. With
${\bm P}_{h\perp}$ taken into account, new PFF's arise. Using the
Lorentz- and P-invariance one can write in the leading twist
approximation 8 independent spin structures \cite{muldt,muldz}. Most
spectacularly it is seen in the helicity basis where one can build 8
twist-2 combinations, linear in spin matrices of the quark and hadron
{$\bm\sigma$}, ${\bm S}$ with momenta ${\bm k}$,
${\bm P}_h$.

Especially interesting is a new chiral-odd  structure that describes
a left--right asymmetry in the fragmentation of a transversely
polarized quark:
$$
H_1^\perp\mbox{\bm\sigma}({\bm k}\times
{\bm P}_{h\perp})/k\la P_{h\perp}\ra,
$$
where $H_1^\perp$ is a
function of the longitudinal momentum fraction $z$ and quark
transverse momentum  $k_T^2$. The $\la P_{h\perp}\ra$ is the averaged
transverse momentum of the final hadron \footnote{We use the notation
of the work \cite{muldt,muldz,mulddis}.  Notice different
normalization factor, $\la P_{h\perp}\ra$ instead of $M_h$.}.  Since
the $H_1^\perp$ term is chiral-odd, it makes possible to measure the
proton transversity distribution $h_1$ in semi-inclusive DIS from a
transversely polarized target by measuring the left-right asymmetry
of forward produced pions (see~\cite{mulddis,kotz} and references
therein). It serves as analyzing power of the Collins effect.

The problem is that, first, this function was completely unknown
till recent time both
theoretically and experimentally. Second, the function $H_1^\perp$
is the so-called T-odd fragmentation function: under the naive time
reversal ${\bm P},\ {\bm k},\ {\bm S}$ and $\bm\sigma$
change sign, which demands a purely imaginary (or zero) $H_1^\perp$
in the contradiction with naive hermiticity. This, however, does not mean
the break of T-invariance but rather the presence of an interference
of different channels in forming the final state with different
phase shifts, like in the case of single spin asymmetry
phenomena~\cite{gasior}. A simple model for this function could be
found in~\cite{collins}
\footnote{It was however
conjectured~\cite{jjt} that the final state phase shift can be
averaged to zero for a single hadron fragmentation upon summing over
unobserved states $X$.}.
(In this aspect they are very different with the T-odd PDF's which
can not exist since they are purely real.  Interaction among initial
hadrons which could brings an imaginary part breaks the factorization
and the whole parton picture.) Thus, the situation here is far from
being clear.

Meanwhile, the data collected by DELPHI (and other LEP experiments)
give a possibility to measure $H_1^\perp$.  The point is that despite
the fact that the transverse polarization of quarks in Z$^0$ decay is
very small ($O(m_q/M_Z)$), there is a non-trivial correlation between
transverse spins of a quark and an antiquark in the Standard Model:
$C^{q\bar q}_{TT}= {(v_q^2-a_q^2)/(v_q^2+a_q^2)}$, which are at $Z^0$
peak: $C_{TT}^{u,c}\approx -0.74$ and $C_{TT}^{d,s,b}\approx -0.35$.
With the production cross section ratio $\sigma_u/\sigma_d=0.78$ this
gives for the average over flavors value $\la C_{TT}\ra\approx -0.5$.

The transverse spin correlation results in a peculiar azimuthal angle
dependence of produced hadrons, if the T-odd fragmentation function
$H_1^\perp$ does exist~\cite{collins,colpsu}.  A simpler method has
been proposed by an Amsterdam group \cite{muldz}. They
predict a specific azimuthal asymmetry of a hadron in a jet around
the axis in direction of a second hadron in the opposite jet
\footnote{The factorized Gaussian form of $P_{h\perp}$ dependence was
assumed for $H_1^{q\perp}$ and $D_1^q$ and integrated over
$|P_{h\perp}|$.}

\begin{eqnarray}
{{\rm d}\sigma\over {\rm
d}\cos\theta_2 {\rm d}\phi_1}\propto (1+\cos^2\theta_2)\cdot \left(1+
{6\over\pi}\left[{H_1^{\perp q}\over D_1^q}\right]^2 C_{TT}^{q\bar
q}{\sin^2\theta_2\over 1+\cos^2\theta_2}\cos(2\phi_1)\right) \, ,
\label{mulders}
\end{eqnarray}
where $\theta_2$ is the polar angle of the electron and the second
hadron momenta ${\bm P}_2$, and $\phi_1$ is the azimuthal angle
counted off the $({\bm P}_2,\, {\bm e}^-)$-plane.
This asymmetry was recently measured \cite{todd} using the DELPHI data
collection.  For the leading charged particles (mostly pions) in each
jet of two-jet events, summed over $z$ and averaged over quark
flavors (assuming $H_1^{\perp}=\sum_H H_1^{\perp\, q/H}$ is flavor
independent), the most reliable preliminary value of the analyzing
power is found to be

\begin{equation}
\left|{\la H_1^{\perp}\ra\over\la D_1\ra}\right| =(6.3\pm 2.0)\% \, ,
\label{apower}
\end{equation}
with presumably large systematic errors \footnote{Close value
was also obtained from pion asymmetry in inclusive $pp$-scattering
\cite{bogl99}.}.

\section{Azimuthal asymmetries}

The azimuthal asymmetries measured by HERMES for SIDIS with $\pi^+$
and $\pi^-$ production on longitudinally polarized target are

\be
A_{UL}^W=2\frac{\int d\phi dyW \left(dN^+/S^+dyd\phi
-dN^-/S^-dyd\phi\right)}
{\int d\phi dy\left(dN^+/dyd\phi
+dN^-/dyd\phi\right)}\ ,
\label{asymw}
\ee
where $W=\sin\phi$ or $\sin2\phi$ and $S^\pm_H$ is the nucleon
polarization. It consists of two sorts  terms \cite{mulddis}: a
twist-2 asymmetry $\sin2\phi$
\be
A_{UL}^{\sin2\phi}\propto -
\frac{\sum_a e_a^2h^{\perp(1)a}_{1L}(x)\la H^{\perp a/\pi}_1(z)\ra}
{\sum_a e_a^2 f^a_1(x)\la D_1^{a/\pi}(z)\ra}\,,
\label{atwist2}
\ee
and a twist-3 asymmetry $\sin\phi$.
\be
A_{UL}^{\sin\phi}\propto\frac{8M}{Q}\cdot
\frac{\sum_a e_a^2\biggl( xh^a_L(x)\la H^{\perp a/\pi}_1(z)/z\ra
-h^{\perp(1)a}_{1L}(x)\la\widetilde H^{a/\pi}(z)/z\ra\biggr)}
{\sum_a e_a^2 f^a_1(x)\la D_1^{a/\pi}(z)\ra}\,.
\label{atwist3}
\ee
Here $\phi$ is the azimuthal angle around the $z$-axis opposite to
direction of virtual $\gamma$ momentum in the Lab frame, counted from
the electron scattering plane.
The first asymmetry is proportional to the $p_T$-dependent transverse
quark spin distribution in longitudinally polarized proton,
$h_{1L}^\perp(x,p_T)$, while the second one contains two parts:  one
term is proportional to the twist-3 distribution function  $h_L(x)$
and the second one proportional to the twist-3 interaction dependent
correction to the fragmentation function $\widetilde H$. We will
systematically disregard this interaction dependent correction
(just as in the Wandzura-Wilczek relation)
\footnote{The calculations in the
instanton model of QCD vacuum supports disregard of $\widetilde h_L$
\cite{DPG99}. As for $\widetilde H$, it disappears after integration
over $z$ due to relation \cite{muld00} ${\widetilde H}\propto
z{d\over dz}(zH_1^\perp(z))$.}.

In the same approximation, the integrated functions over the quark
transverse momentum, $h_{1L}^\perp(x,p_T)$ and $h_L(x)$, are
expressed through the transversity $h_1$
\be
h_{1L}^{\perp(1)}(x)\equiv\int d^2p_T\left(\frac{p_T^2}{2M^2}\right)
h_{1L}^{\perp}(x,p_T)=-x^2\int_x^1d\xi h_1(\xi)/\xi^2=-(x/2)h_L(x) \, ,
\label{wwform}
\ee

Assume now that only the favored fragmentation functions
$D_1^{a/\pi}$ and $H_1^{\perp a/\pi}$ will contribute this ratios,
i.e.  $D_1^{u/\pi^+}(z)=D_1^{\bar d/\pi^+}(z)=D_1^{d/\pi^-}(z)
=D_1^{\bar u/\pi^-}(z)\equiv D_1(z)$ and similarly for
$H^{\perp}_1(z)$. This would allow us to extract from the observed
HERMES asymmetries an information on $h_1^u(x)+(1/4)h_1^{\bar d}(x)$
and to compare with some model prediction. Instead, we use the
prediction of the chiral soliton model \cite{pp96} for $h_1^a(x)$ and
the GRV parametrization \cite{GRV} for unpolarized PDF's $f_1^a(x)$
to calculate the asymmetries $A_{UL}^{\sin\phi}$ and
$A_{UL}^{\sin2\phi}$ for $\pi^+$ and $\pi^-$. The comparison of the
asymmetries thus obtained with the  HERMES experimental data is
presented on Fig. 2.

\begin{figure}[ht]
\begin{center}
\raisebox{-40mm}{
\mbox{\epsfig{figure=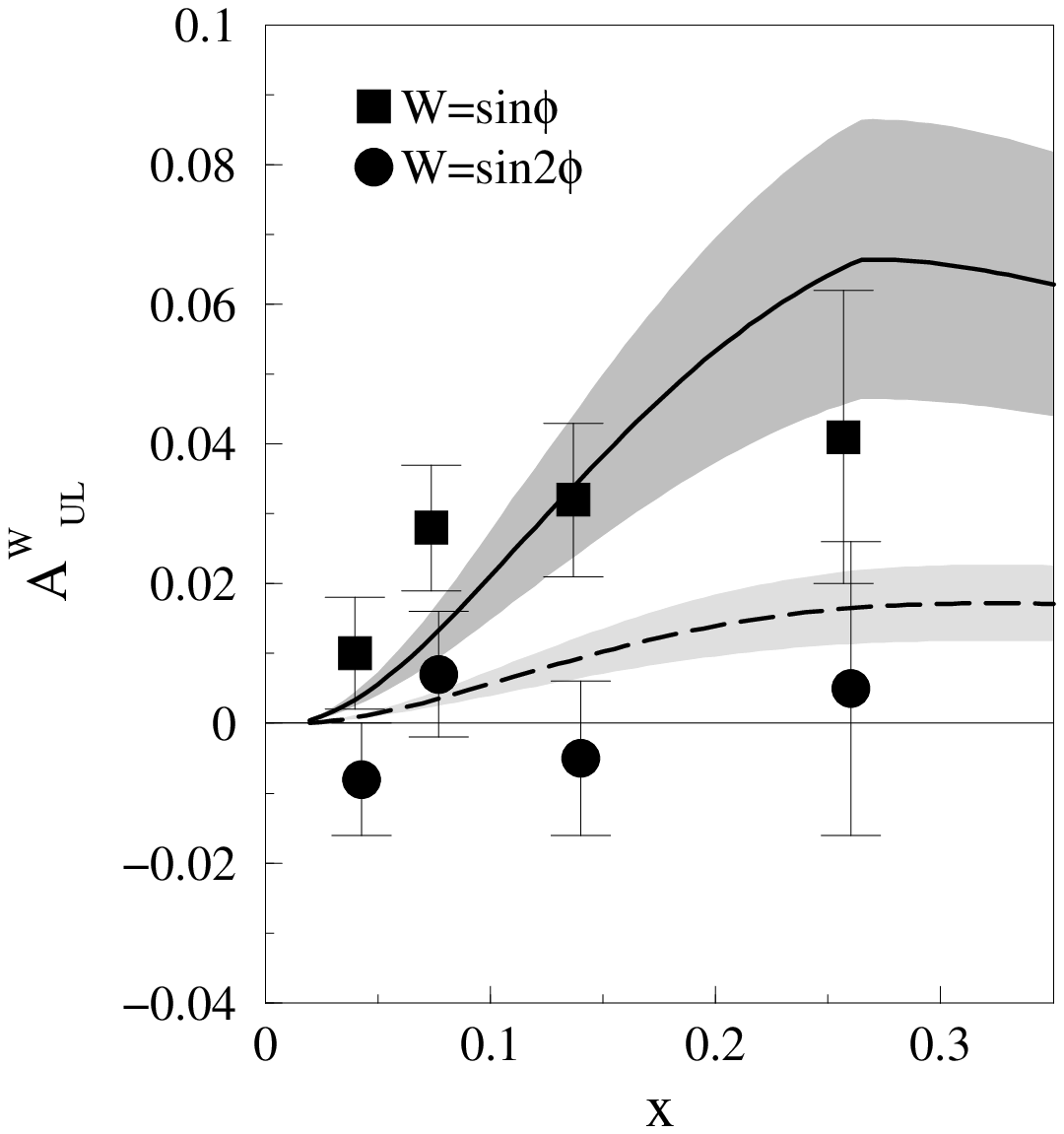,width=6.0cm,height=6.0cm}}}
\raisebox{-40mm}{
\mbox{\epsfig{figure=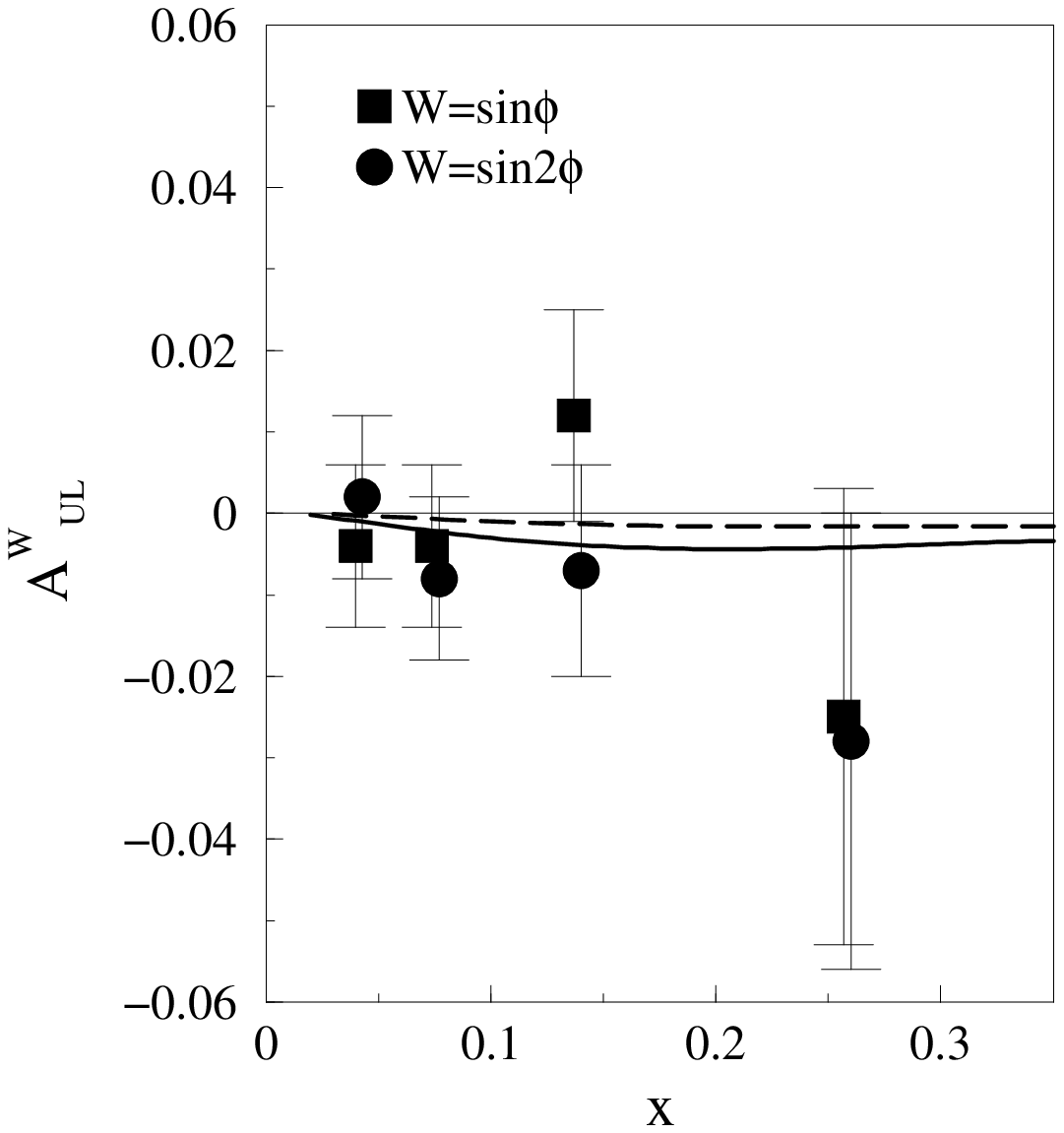,width=6.0cm,height=6.0cm}}}
\end{center}
{\footnotesize
{\bf Fig. 2.} Single spin azimuthal asymmetry for
$\pi^{+}$ ({\bf left}) and $\pi^{-}$ ({\bf right}):
$A_{UL}^{\sin\phi}$ (squares) and $A_{UL}^{\sin2\phi}$ (circles) as a
functions of x. The solid ($A_{UL}^{\sin\phi}$) and the dashed lines
($A_{UL}^{\sin2\phi}$) correspond to the chiral quark-soliton
model calculation at
$Q^2=4\,GeV^2$. The shaded areas represent the
uncertainty in the value of the analyzing power (\ref{apower}).}
\end{figure}

The agreement is good enough though the experimental errors are yet
rather large. The sign of the asymmetry is uncertain since only the
modulus of the analyzing power (\ref{apower}) is known.  Fig. 2 gives
evidence for positive sign. Notice that in spite of the factor
${M}/{Q}$ in the Exp.  (\ref{atwist3}) it is several times larger
than (\ref{atwist2}) for moderate $Q^2$. That is why this asymmetry
prevails for the HERMES data where  $\la Q^2\ra\approx 2.5\,GeV^2$.
One can thus state that the effective chiral quark soliton model
\cite{pp96} gives a rather realistic picture of the proton
transversity $h_1^a(x)$.

The interesting observable related to $h_1(x)$ is the proton tensor
charge. The  calculation in this model  yields \cite{pol96}
$h_1\equiv \sum_a\int_0^1 dx\Big(h_1^a(x)-h_1^{\bar a}(x)\Big)=0.6$.
Compare this with most recent experimental value of
the proton axial charge \cite{leader99} $a_0=0.28\pm0.05$ and the
value obtained in the same model $a_0 = 0.35$ \cite{anomaly}. These
very different values for the axial and tensor charges of the
nucleon are in contradiction with the nonrelativistic quark model
prediction.

Concerning the asymmetry observed by SMC \cite{bravardis99} on
transversely  polarized target one can state that it agrees with
result of HERMES (see \cite{egpu00} for more details).

\bigskip
In conclusion, using the preliminary estimation for Collins effect
analyzing power from DELPHI data and the effective chiral quark
soliton model for the proton transversity distribution we obtain a
rather good description of the azimuthal asymmetries in
semi-inclusive hadron production measured by HERMES and SMC, though
the experimental errors are yet large. This, however, is only the
first experiments!  We would like to stress that our description has
no free adjustable parameters.  Probably the most useful lesson we
have learned is that to measure transversity in SIDIS in the region
of moderate $Q^2$ it is not necessary to use a transversely polarized
target. Due to approximate Wandzura-Wilczek type relations
(\ref{wwform}) one can explore the longitudinally polarized target
also.  This is very important for future experiments, like COMPASS at
CERN since the proton transversity could be measured {\bf
simultaneously} with the spin gluon distribution $\Delta G(x)$.



\end{document}